\documentclass[aps,pra,twocolumn,groupedaddress,nofootinbib]{revtex4-1}

\usepackage{graphicx}
\usepackage{amsfonts}
\usepackage{amsmath}
\usepackage{relsize}
\usepackage{color,soul}
\usepackage{textcomp}
\usepackage{tikz}
\usepackage{pgfplots}
\usepackage[bottom]{footmisc}
\usepackage{cleveref}
\usepackage{soul}
\usepackage[normalem]{ulem}
\pgfplotsset{compat=1.3}
\pgfplotsset{colormap={jet}{rgb255(0)=(0,0,143);rgb255(8)=(0,0,255);rgb255(24)=(0,255,255);rgb255(40)=(255,255,0);rgb255(56)=(255,0,0);rgb255(64)=(128,0,0)}}
\usepackage{tikz}
\usepackage{circuitikz}
\usepackage{dsfont}

\newcommand{\ket}[1]{|#1 \rangle}
\newcommand{\bra}[1]{\langle #1|}

\begin{document}

\title{Evaluating the performance of sigmoid quantum perceptrons in quantum neural networks}

\author{Samuel A. Wilkinson}
\affiliation{Friedrich-Alexander University Erlangen-N\"urnberg (FAU), Department of Physics, Erlangen, Germany}
\author{Michael J. Hartmann}
\affiliation{Friedrich-Alexander University Erlangen-N\"urnberg (FAU), Department of Physics, Erlangen, Germany}

\date{\today}

\begin{abstract}
Quantum neural networks (QNN) have been proposed as a promising architecture for quantum machine learning. There exist a number of different quantum circuit designs being branded as QNNs, however no clear candidate has presented itself as more suitable than the others. Rather, the search for a ``quantum perceptron" -- the fundamental building block of a QNN -- is still underway.
One candidate is quantum perceptrons designed to emulate the nonlinear activation functions of classical perceptrons. Such sigmoid quantum perceptrons (SQPs) inherit the universal approximation property that guarantees that classical neural networks can approximate any function. However, this does not guarantee that QNNs built from SQPs will have any quantum advantage over their classical counterparts. Here we critically investigate both the capabilities and performance of SQP networks by computing their effective dimension and effective capacity, as well as examining their performance on real learning problems. The results are compared to those obtained for other candidate networks which lack activation functions. It is found that simpler, and apparently easier-to-implement parametric quantum circuits actually perform better than SQPs. This indicates that the universal approximation theorem, which a cornerstone of the theory of classical neural networks, is not a relevant criterion for QNNs.
\end{abstract}

\maketitle

\section*{Introduction}

In recent years there has been a proliferation of proposals for quantum algorithms and circuits dubbed ``quantum neural networks'' (QNNs) \cite{Schuld2014,Cong2019,Beer2020,Herrmann2021}. Given the immense success of artificial neural networks in classical computing, it is not surprising that a quantum analogue would be greatly sought after. In a classical setting, neural networks provide a powerful ansatz for an enormous range of machine learning tasks, in many cases performing tasks that present insurmountable obstacles for traditional computing \cite{Goodfellow2016}. Thus, to develop quantum algorithms comparable to the classical cutting edge, it is natural to try to adopt neural networks to a quantum setting. However, some basic problems immediately present themselves: classical neural networks gain much of their expressive power from nonlinear activation functions and dissipative dynamics, whereas quantum mechanics is manifestly linear and the unitary evolution provided by quantum gates is reversible and non-dissipative. Several different approaches have been employed to circumvent these issues and construct a quantum neural network. The most common has been to simply ignore it: construct parameterised quantum circuits which consist only of linear, unitary operations train these to perform learning tasks. These circuits use neither the layered structure of a neural network nor the nonlinear activation functions, and thus bare very little resemblance to either the artificial neural networks used in classical computing, nor the biological neural networks in living organism which inspired them. Nevertheless, the approach of using simple parameterised quantum circuits is attractive for a number of reasons, and they have proven capable of a number of learning tasks \cite{Romero2017, Farhi2018, Havlicek2019}.

Another approach would be try to engineer a nonlinear activation function in quantum hardware, thus making a quantum circuit more ``neural." There have been several different approaches used \cite{Kak1995,Tacchino2019,Niu2020, Yan2020}. Here we will restrict our attention to situations where an operation on a quantum circuit implements a map such that the excitation probability of the target qubit is a nonlinear function of the state of an input or control qubit (as well as of some tunable parameters). Thus, we consider a map of the form
\begin{equation} \label{eq:SQPmap}
	\begin{split}
	\ket{\textbf{x}_\textrm{in}}\ket{0_\textrm{target}} \rightarrow \ket{\textbf{x}_\textrm{in}} &\left[ \sqrt{1 - f(\textbf{a}_\textrm{in})}\ket{0_\textrm{target}} \right. \\ 
	& + \left. \sqrt{f(\textbf{a}_\textrm{in})}\ket{1_\textrm{target}} \right],
	\end{split}
\end{equation}
where $\textbf{a}_\textrm{in} = \sum_{j\in\textrm{in}} w_j \hat{x}_\textrm{in} - b$ and $f(x)$ is a non-linear sigmoid activation function. The weights $w_j$ and bias $b$ are parameters which can be tuned variationally.
We will adapt the more general term ``sigmoid quantum perceptron'' (SQP) to refer to any operation on a quantum circuit of this form.

The SQP is motivated by the design of classical sigmoid perceptrons, and in particular by the universal approximation theorem (UAT) \cite{Cybenko1989, Hornik1989} which states that a network of perceptrons with sigmoid activation functions can approximate any smooth function arbitrarily well with just a single hidden layer. The UAT guarantees that a neural network with sigmoid perceptrons will be highly expressive, and this high expressiveness is one of the reasons neural networks have been so successful in such a range of classical machine learning applications.

As we will discuss below, SQPs inherit the universal approximation property. A SQP is able to emulate the behaviour of a single classical perceptron, which indicates that a learning model generated from a network of SQPs should be able to emulate the behaviour of a classical neural network, at least up to a single layer.
Further, it has been argued that strictly quantum properties of the SQP should give it expressive power above and beyond that which is attainable by classical perceptrons.

These preliminary results are suggestive, but not decisive. Here, we examine thoroughly and critically the actual performance and expressive power of networks of SQPs, compared with simpler variational quantum circuits.
We find that, despite the promised universality, SQPs are \textit{less} expressive than more conventional variational circuits according to every metric we study.
In particular, we find that SQPs strongly mimic CNNs, and as such they can perform essentially classical tasks like classification of real-world data, but they perform very poorly on tasks involving quantum data (where the inputs and outputs are quantum states).
Thus, while SQPs achieve the originally stated goal of implementing classically-inspired nonlinear activation functions, we find that this does not give them any advantage over more conventional, hardware efficient quantum circuits.

Understanding, increasing and controlling the expressive power of QNNs is an important task.
While there has beenwork in the past comparing the expressiveness of QNNs with that of CNNs \cite{Abbas2021,Qian2021}, there has recently been a growing body of research comparing the expressiveness of different QNNs, in an attempt to pin down appropriate and effective circuit ansätze \cite{Beer2021,Funcke2021}.
Here we investigate, through a series of numerical simulations, the impact that choice of gates and choice of network structure has on the ability of a network to perform learning tasks. 
Other work has shown that the expressiveness of a network can be increased by using multi-qubit potentials \cite{Ban2021}, by choice of quantum feature map \cite{Havlicek2019} and by repeatedly re-uploading input data \cite{Schuld2021}.

The paper is structured as follows: first, in Section~\ref{sec:Perceptron} we introduce the SQP in detail. Much of this material is already covered in \cite{Torrontegui2019}, but we repeat it here for completeness. We also introduce the other quantum circuits against which the SQP will be compared. Then, in Section~\ref{sec:Expressive} we consider different methods of measuring and quantifying the expressive power of a network, and compare the expressive power of a SQP network against that of more standard variational quantum circuits. In Section~\ref{sec:Experiments} we examine how well SQP networks performs at classification tasks involving real-world data sets, where again we compare the performance against that of other variational quantum circuits. Throughout, we will be interested in the effect that gate choice, circuit depth and choice off data encoding have on the network's expressiveness, trainability and generalisability.

\section{Quantum neural networks}\label{sec:Perceptron}

\subsection{The universal approximation property in classical neural networks}

A classical perceptron calculates the activation of a neuron by taking the weighted sum of neurons in the previous layer and feeding this sum into a nonlinear function, taking the output of this function as the activation of the neuron,
\begin{equation}\label{eq:activation}
	a_k = f\left( \sum_{j<k} w_{jk} a_j - b_k \right)
\end{equation}
where the weights $w_{jk}$ and the bias $b_k$ are tunable parameters.
These perceptrons form the basic building blocks of feedforward networks, where the activation of each layer is determined by the activation of the previous layer in a highly nonlinear fashion. 
The nonlinear function $f$ is called the ``activation function'', as it determines whether the neuron is ``activated'' or not.
These activation functions are typically chosen to be of a form that satisfies the universal approximation theorem (UAT).

Loosely speaking, the UAT states that a neural network endowed with a sigmoid activation function can approximate any continuous function with just a single hidden layer (albeit of arbitrarily large depth). More precisely, let $I_N = [0,1]^N$ be the $N$-dimensional unit cube and let $C(I_N)$ be the space of continuous functions on $I_N$. Further, let $f$ be a continous sigmoid function -- that is, a continuous function such that $f(\infty) \rightarrow 1$ and $f(-\infty) \rightarrow 0$. Then, sums of the form
\begin{equation}\label{eq:UAT sum}
	g(\mathbf{x}) = \sum_{j}^{M} \alpha_j f\left( \sum_{k=1}^{N} w_{jk}x_k - b_j \right)
\end{equation}
with finite $M$ are dense in $C(I_N)$, and where $x_k$ are the elements of $\mathbf{x}$. This means that for any given target function $G\in C(I_N)$ and for any desired accuracy $\epsilon$ there exists a finite sum of the above form such that $|G(\textbf{x}) - g(\textbf{x})| \leq \epsilon$ for all $\textbf{x}\in I_N$.

This property guarantees the expressive power of CNNs. However, for CNN to have this property, it is required that the activation function be non-linear\footnote{The theorem as stated above requires a sigmoid activation function. More general versions of the UAT allow for other nonlinear activation, such as the popular rectified linear unit (ReLU) activation.}. For example, a neural network without an activation function (i.e. a linear network) is completely unable to express an XOR function. With a sigmoid activation function, however, learning XOR is a trivial matter.

\subsection{QNNs with classical and quantum data}

While a CNN generates a statistical model $f_\theta(\textbf{x})$ via a nested series of weighted sums and nonlinear activations of the form Eq.~\ref{eq:activation}, a QNN generates a statistical model by first loading classical input data into a register of qubits, then applying a series of unitary operations on the qubits, before finally measuring the expectation value of some observable.

The bulk of a QNN consists of a sequence of parameterised unitary operations,\footnote{Non-parameterised gates may also play an important role, but for our current purposes we can simply absorb them into our parameterised unitaries.}
\begin{equation}
	\mathcal{U}(\vec{\theta}) = \prod_j^{N_P} U_j(\theta_j).
\end{equation}
A vector of classical input data $\mathbf{x}$ may be input into the network in the form of a quantum feature map $F(\mathbf{x})$, preparing the state
\begin{equation} \label{eq:outputDensity}
	\rho_\mathbf{x} = \left[F(\mathbf{x})\ket{0}_\textrm{in}\bra{0}_\textrm{in}  F^\dagger(\mathbf{x})\right] \otimes \ket{0}_{\bar{\textrm{in}}}\bra{0}_{\bar{\textrm{in}}}
\end{equation}
where the subscript ``in'' indicates the input qubits and the subscript``$\bar{\textrm{in}}$'' indicates any qubits not in the input layer (thus, all qubits in hidden and output layers. For the HEA, there are no qubits in ``$\bar{\textrm{in}}$'').
Network predictions can then be read out in the form of the expectation values of some observables $O_\alpha$,
\begin{equation}
	h_{\theta,\alpha}(\mathbf{x}) = \textrm{Tr}\left[ \mathcal{U}(\theta)\rho_\mathbf{x}\mathcal{U}^\dagger(\theta) O_\alpha \right].
\end{equation}
For compactness of notation, we will typically drop the subscript $\alpha$.
$h_\theta(\mathbf{x})$ is referred to as the statistical model (or ``hypothesis'') generated by our QNN.

`Training' the network consists of finding parameters $\vec{\theta}$ such that our model $h_\theta(\textbf{x})$ closely approximates some target function $y(\textbf{x})$.
We will be concerned here with supervised learning, as that is conceptually the simplest kind of learning algorithm, and thus we will assume that we have labelled training data at some certain values of $\textbf{x}$, that is there exists a set of points $j$ such that we already know $y(\textbf{x}_j)$.
The possibility of noise in the training labels and in the QNN itself are both neglected for the present work.

Thus, a QNN consists of a feature map $F(\textbf{x})$ via which we load input data into the network, a set of parameterised unitaries $U(\theta)$ which processes the data, and finally a set of observables $O_\alpha$ which are measured at the end of the circuit.
In the present work, we are concerned with the effect of changing the choice of parameterised unitaries, and thus when we refer to different QNNs we generally mean different choices of gates in $U(\theta)$.

One could also consider a situation where the input to the network is not classical data, but rather a quantum state. In such a situation, there is no need for a feature map, as the state can be processed on the quantum computer directly. Further, we may consider a situation where, instead of obtaining a probability distribution at the end of the network, we wish instead to output a quantum state. Thus, we need not specify observables $O_\alpha$, but rather take the final state of the output qubits to be our output.
This is the situation considered in Section~\ref{sec:UnitaryLearning}.

\subsection{Sigmoid quantum perceptrons and the universal approximation property for QNNs}

Inspired by the universal approximation capabilities of CNNs, there have been proposals to implement similar non-linear activation functions on quantum hardware.
The idea is to engineer a quantum gate that implements the map given in Eq.~\ref{eq:SQPmap}, so that the excitation probability of the target qubit is given by some sigmoid function $f$. If we require that this map is unitary, then it is given by an operator of the form
\begin{equation} \label{eq:Unitary}
	U = \sum_{\mathbf{x}\in\{0,1\}^{N_\textrm{in}}} \ket{\mathbf{x}}\bra{\mathbf{x}} \otimes V(\mathbf{x}),
\end{equation}
where for each input computational basis state, labelled by a bit-string  $\mathbf{x}$, we apply a single-qubit unitary operator $V(\mathbf{x})$ to the target qubit in the next layer.
For the map to be both unitary and equivalent to Eq.~\ref{eq:SQPmap}, there are two possible matrix forms for $V(\mathbf{x})$:
\begin{equation} \label{eq:Vx}
	V^\pm(\mathbf{x}) = \begin{pmatrix}
		\sqrt{1 - f(a_\mathbf{x})} & \sqrt{f(a_\mathbf{x})}\\
		\pm \sqrt{f(a_\mathbf{x})} & \mp \sqrt{1 - f(a_\mathbf{x})}.
	\end{pmatrix}
\end{equation}
where $a_\mathbf{x}$ is the activation $a_\mathbf{x} = \sum_{j} w_j x_j - b$ and $f(a_\textbf{x})$ is our sigmoid activation function.
We will consider the version with the upper symbols, as this is the unitary given by the protocol of \cite{Torrontegui2019} (a derivation of this specific form is given in Appendix~\ref{app:ARunitary}).
While different hardware implementations lead to slightly different activation functions, we again will consider specifically the form given by the proposal of \cite{Torrontegui2019},
\begin{equation}
	f(x) = \frac{1}{2}\left(1 + \frac{x}{\sqrt{1 + x^2}}\right).
\end{equation}

Treating the SQP as a single unitary operator in this fashion allows us to consider it as a special kind of parameterised quantum gate.
Since the sigmoid function $f(a_\textbf{x})$ is a real-valued function with range $[0,1]$, we can define angles $\theta_\textbf{x}$ such that  $\sqrt{f(a_\mathbf{x})} = \sin\theta_\mathbf{x}/2$ and $\sqrt{1 - f(a_\mathbf{x})} = \cos\theta_\mathbf{x}/2$ and express $V(x)$ as
\begin{equation}
	V(\mathbf{x}) = \begin{pmatrix}
		\cos\theta_\mathbf{x}/2 & \sin\theta_\mathbf{x}/2\\ \sin\theta_\mathbf{x}/2 & -\cos\theta_\mathbf{x}/2
	\end{pmatrix} = ZR_y(\theta_\mathbf{x}).
\end{equation}
From this form, we can see that the SQP can be thought of as a special case of a controlled rotation, where we rotate the target qubit by a different angle for each computational basis state of the input qubits.

Torrentegui and Garcia-Ripoll \cite{Torrontegui2019} have shown that a gate performing such a map will enjoy a quantum version of the UAT. Specifically, they show that any bounded, continuous function $G(\hat{\sigma}_1,\dots,\hat{\sigma}_2) \in [-1,1]$ of the quantum observables $\{\hat{\sigma}_j\}_{j=1}^N$ can be reconstructed up to an error $\epsilon$ onto the state of a qubit using $N$ input qubits and $M+1$ applications of the perceptron gate (with $M$ some finite number of neurons as in Eq.~\ref{eq:UAT sum}). The crux of their argument is that the excitation probability of the target qubit can be written in terms of the $\hat{\sigma}$ observables of the input qubits as a sum of the form Eq.~\ref{eq:UAT sum}. Thus, so long as the function $f$ appearing in Eq.~\ref{eq:SQPmap} is a sigmoid function, a model $h_\theta(\mathbf{x})$ generated by a network of SQPs inherits the universal approximation property enjoyed by classical sigmoid perceptrons.

There have been two major proposals of quantum gates of this kind: the repeat-until-success quantum neuron \cite{Cao2017} which uses a measurement feedback loop to engineer the sigmoid activation\footnote{Despite involving measurements, the repeat-until-success quantum neuron can still be reduced to an effective unitary gate by considering only the input-output relations in the case of a ''successful'' measurement. One recovers a unitary given by Eq.~\ref{eq:Unitary}, where $V(\textbf{x})$ takes the lower symbols of Eq.~\ref{eq:Vx}.}, and the adiabatic ramp quantum neuron \cite{Torrontegui2019} which uses tunable Ising-like interactions. The adiabatic SQP has been experimentally realised in both superconducting \cite{Pechal2021} and ion-trap \cite{Huber2021} qubits.

A difficulty in working with SQPs is that the gate cannot be simply written in the form $e^{-iG\theta}$ for some Hermitian generator G and some real parameter $\theta$. Rather, when written as the exponential as a generator, the gate must take the form $e^{-iG(\boldsymbol{\theta})}$, where the generator is a function of the parameters, involving more than one element of the parameter vector $\boldsymbol{\theta}$.
This means that standard tricks for obtaining gradients such as the parameter shift rule \cite{Mitari2018,Schuld2019,Wierichs2022} cannot be applied to this gate. 
For that reason, in all results presented in this paper gradients are obtained using finite differences. 
(A difference of $\epsilon = 10^{-3}$ is used throughout.)

While the UAT and its quantum counterpart imply that a network of SQP should be able to approximate any continuous function of the observables of the input qubits, given enough qubits and gates, it does not imply that SQP are universal in the sense of forming a universal quantum gate. Indeed, as was shown in \cite{Pechal2021} the SQP can be thought of as a kind of generalised multi-conditional rotation about a \textit{single} axis. It is only when supplemented with general single-qubit rotations that the SQP becomes expressive enough to explore the entire space of unitary operations. When combined with single-qubit rotations, it is possible to implement a CNOT gate with a SQP. 
Since a CNOT and single qubit rotations form a universal gate set, this implies that the SQP and single qubit rotations also form a universal gate set. 
However, such a statement is also true for most two-qubit gates, and alone is not enough to imply that the SQP is especially expressive.

It is also unclear what potential the SQP has for quantum advantage. 
The quantum UAT demonstrates that the SQP has the ability to emulate a classical sigmoid perceptron, but says nothing about the abilities of this gate above and beyond the classical.
Furthermore, since similar universal approximation properties have been shown in single-qubit scenarios \cite{Perez-Salinas2021, Goto2021}, where there is no possibility for entanglement and thus no possibility for quantum advantage.

\subsection{Other networks: Hardware efficient ansatz and Hannover} \label{sec:OtherNetworks}
In the following sections, we pit SQP networks against other quantum neural networks to evaluate their performance. In particular, we compare the SQP network to a hardware efficient ansatz (HEA) and the QNN proposed in \cite{Beer2020}, which we refer to as a Hannover network\footnote{This latter QNN is often referred to as a dissipative QNN due to it's layered, feed-forward structure, where previous layers are traced out, leading to potentially dissipative evolution. However, this structure is exactly identical to that of the SQP network, so we instead refer to this network as the Hannover network, after the city in which it was originally developed.}.

\begin{center}
	\begin{figure}[htb!]
		\includegraphics[scale=1, width=\columnwidth]{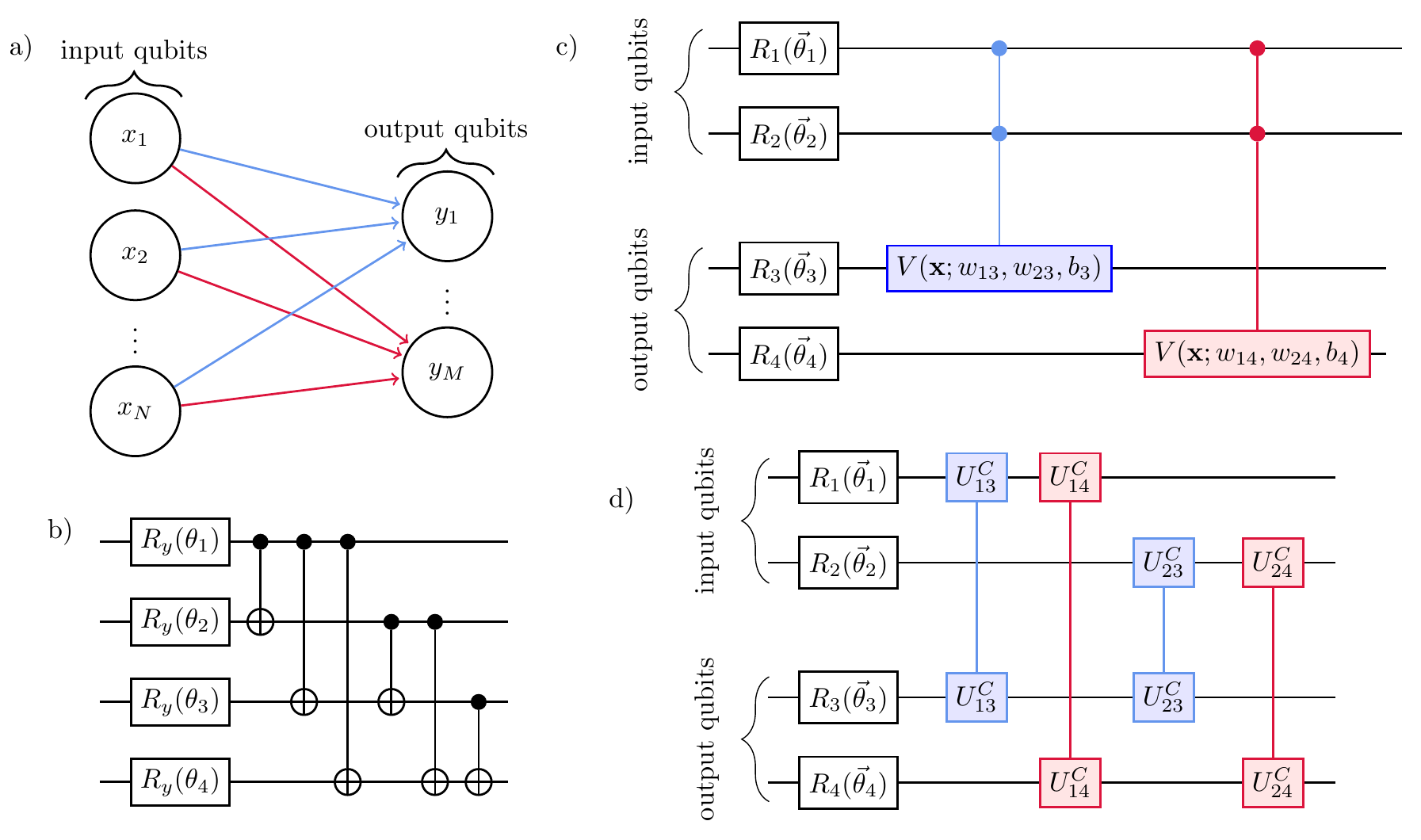}
		\centering
		\caption{\label{fig:Models} a) The general structure of a feedforward network. For a QNN, the circles correspond to qubits and the arrows correspond to many-body quantum gates. b) The basic unit of a hardware efficient ansatz (HEA) QNN. This circuit corresponds to a single layer of the network. Applying this circuit in sequence, but with new parameters in the $R_y(\theta)$ gates, creates a multi-layer network. The structure of this network is fundamentally different from that of a feedforward neural network, as all layers involve the same qubits. c) The basic unit of a sigmoid quantum perceptron (SQP) network. This mimics the structure of a feedforward network, where SQPs provide interactions between layers, feeding the input data forward. d) The basic unit of the Hannover network, which is similar to the SQP network but with inter-layer interactions mediated by the ''canonical'' gate, rather than a SQP. The Hannover network also has a feedforward structure.}
	\end{figure}
\end{center}

The Hannover network is chosen as a network with an identical network topology to the SQP network, but with a different choice of multi-qubit gates. Whereas the SQP networks features SQPs between layers, the Hannover network uses the so-called canonical gate
\begin{equation} \label{eq:CANgate}
	U_{ij}^\textrm{CAN}(\theta_x,\theta_y,\theta_z) = \exp\left[-\frac{i}{2}\left(\theta_x X_i X_j + \theta_y Y_i Y_j + \theta_z Z_i Z_j \right)\right]
\end{equation}
where $X$, $Y$ and $Z$ are the Pauli operators and the subscripts indicate which qubits they act on.
The network is designed to feed information on to other layers, imitating a feedforward network.
Thus, the indicies $i$ and $j$ in Eq.~\ref{eq:CANgate} always refer to qubits in two different layers, as can be see in Fig.~\ref{fig:Models}.

The hardware efficient ansatz is chosen to be a parameterised quantum circuit which is easy to implement on current and near-term hardware. It consists only of parameterised single-qubit $Y$-rotations, $R_y(\theta) = e^{-i\frac{\theta}{2} Y}$, and CNOTs to act as entangling gates. This network has a layered structure, where each layers contains a single qubit rotation on each qubit followed a CNOT between each pair of qubits in the circuit\footnote{In certain hardware implementations a CZ gate is more convenient to implement than a CNOT gate. This does not alter any of our results significantly.}.

We wish to emphasise that there are two different network topologies at play here. The SQP and Hannover networks employ a feedforward structure whereby different layers of the network correspond to different qubits. After the two-qubit gates are applied, feeding data from one layer to the next, the previous layer is discarded -- mathematically, it is traced out. On the HEA circuit, however, the same qubits are used for every layer, so that output data is read out from the very same qubits input data was loaded into.

\onecolumngrid

\twocolumngrid

\section{Measuring and quantifying expressive power of a network}\label{sec:Expressive}

The expressive capacity of a learning model -- that is, it's ability to fit various different functions -- can be quantified in many ways. Perhaps the most important from a foundational perspective is the Vapnik-Chervonenkis (VC) dimension, which is defined as the maximum size of a set that can be ``shattered'' by a hypothesis class $\mathcal{H}$ (here we may think of our ansatz circuit as a hypothesis class) \cite{Shalev-Shwarty2014}. The VC dimension appears in many of the most fundamental results in statistical learning theory, including fundamental theorem of PAC (``probably approximately correct") learnability \cite{Blumer1989}, but in many practical instances the VC dimension is extremely difficult to compute.
Thus other, more practical measures have been proposed, which heuristically capture notions of the expressiveness, learnability and ``power'' of a learning model.
We focus on the so-called ``effective dimension," to be defined below, as this quantity has previously been used to argue that parameterised quantum circuits can have greater expressive power than their classical counterparts \cite{Abbas2021}.

In the neural network literature, network designs are often compared by discussing their ``expressiveness'' or their ``capacity''. These terms are sometimes used interchangeably, and indeed for our present purposes the difference between them is largely unimportant. Expressiveness is a measure of the size of a family of models, essentially asking how many different models can be found by a given network. The capacity of a family of models is a measure of the ability of a network to simply fit arbitrary data -- effectively, the ability of the network to act as a memory \cite{Zhang2016}. Below, the effective dimension will act as a measure of the expressiveness of QNNs, while the effective capacity is, unsurprisingly, a measure of the capacity.

\subsection{Effective dimension}\label{sec:EffectiveDimension}

The effective dimension of a statistical model was introduced as a measure of the complexity of a model that accounts for both the intrinsic curvature of the statistical manifold associated with the model and the amount of data available for training \cite{Berezniuk2020}. 
It is then defined in a way analogously to the box-counting dimension of a set. 
The box-counting dimension is defined as the logarithm of the number of boxes needed to cover the set as the size of the boxes goes to zero, 
\begin{equation}
\textrm{dim}_\textrm{box}(S) = \lim_{\epsilon\rightarrow 0} \frac{\log N(\epsilon)}{|\log\epsilon|}.
\end{equation}
Motivated by this, the effective dimension of a parametric family $\mathcal{M}$ on a parameter space $\Theta$ with access to $n$ observations is
\begin{equation}
	\textrm{dim}_{\textrm{eff},n} \approx \frac{\log\left(\textrm{no. unit cubes needed to cover} \sqrt{\frac{n}{2\pi}} \Theta\right)}{|\log\sqrt{\frac{2\pi}{n}}|}.
\end{equation}

To define the notion of a cube (and geometry more broadly) within model space (as opposed to simply Euclidean parameter space), we will here introduce the Fisher information matrix (FIM), which servers as a metric for the space of models.

The FIM is given by
\begin{equation} \label{eq:FisherInfo}
	F_{ij}(\boldsymbol{\theta}) = \mathbb{E} \sum_\alpha \left[ \frac{\partial \log f_{\theta,\alpha}(\mathbf{x}) }{\partial\theta_i} \frac{\partial \log f_{\theta,\alpha}(\mathbf{x}) }{\partial\theta_j}  \right]
\end{equation}
where the expectation value is taken with respect to the input data $\mathbf{x}$.
This can be loosely interpreted as a measure of how easy it is to determine the parameters $\boldsymbol{\theta}$ based on samples from the probability distribution $h_{\boldsymbol{\theta}}(x)$.
Let $\boldsymbol{\phi}$ be our estimator for $\boldsymbol{\theta}$.
Since $\boldsymbol{\phi}$ is constructed from random samples from $h$, it is itself a random vector (its elements are random variables).
The covariance matrix tells us the covariance of the elements of this estimator, $K_{ij}(\boldsymbol{\phi}) = \mathbb{E}[\phi_i\phi_j] - \mathbb{E}[\phi_i]\mathbb{E}[\phi_j]$, and the FIM bounds 
$K(\boldsymbol{\phi})$ by
\begin{equation}
	K(\boldsymbol{\phi}) \geq \frac{1}{N_S} F^{-1}(\boldsymbol{\theta})
\end{equation}
where $N_S$ is the number of samples drawn from $p_\theta$, and the inequality between two matrices $A \geq B$ means that $\mathbf{v}^T A \mathbf{v} \geq \mathbf{v}^T B \mathbf{v}$ for any vector $\mathbf{v}$.

The above formulation of the FIM, by far the most common in the literature, is specifically for a maximum likelihood estimator.
This is the more natural and useful quantity for applications in quantum sensing, and in many machine learning applications.
However, we can also define a Fisher information for a least-squares cost function \cite{Pennington2018a}.
This gives us
\begin{equation}
	F_{ij}^{\textrm{LS}}(\boldsymbol{\theta}) = \mathbb{E} \sum_\alpha \left[ \frac{\partial f_{\theta,\alpha}(\mathbf{x}) }{\partial\theta_i} \frac{\partial f_{\theta,\alpha}(\mathbf{x}) }{\partial\theta_j}  \right].
\end{equation}

The FIM also has a geometric interpretation. Consider a parametric family of models $\mathcal{M}$ such that every particular model $f_\theta$ can be considered an element of $\mathcal{M}$.
We can cover the set $\mathcal{M}$ with a set of coordinates $\Theta$, so that every $\theta \in \Theta$ corresponds to a model $f_\theta \in \mathcal{M}$.
This allows us to define a statistical manifold.
If we wish discuss geometric properties of this manifold, such as the distance between distributions, or the volume of some space, we must impose a metric. 
A natural choice for metric is the FIM (and, indeed, it has been shown that any metric satisfying certain desirable conditions must be the Fisher information up to a scalar factor \cite{Meyer2021}).
In this way, we can interpret the FIM as telling us about the curvature of our model space.

Now that we have a natural notion of geometry in the space of models, we can give an explicit formula for the effective dimension of a parametric family \cite{Berezniuk2020}.
\begin{equation}
	\dim_{\textrm{eff},n}(\mathcal{M}) = 2 \frac{\log\left(\frac{1}{V_\Theta}\int_\Theta\sqrt{\det\left(\mathds{1} + \frac{n}{2\pi} \hat{F}(\theta) \right)} \right)}{\log\frac{n}{2\pi}}.
\end{equation}
$V_\Theta = \int_\Theta d\boldsymbol{\theta}$ is the volume of the parameter space (if one restricts parameters to finite intervals, then one can rescale the parameters to set $V_\Theta=1$).
$\hat{F}$ is the normalised Fisher matrix, defined as
\begin{equation}
	\hat{F}_{ij}(\boldsymbol{\theta}) = N_P \frac{V_\Theta}{\int_\Theta \textrm{Tr}[F(\boldsymbol{\theta})]d\boldsymbol{\theta}} F_{ij}.
\end{equation}
This normalisation ensures that $\frac{1}{V_\Theta}\int_\Theta \textrm{Tr}[\hat{F}(\boldsymbol{\theta})] = N_P$. Finally, $n$ is the number of observations (input training data) available to the model. For numerical calculations, integrals over the parameter space are approximated by a sum over randomly sampled points.

In \cite{Abbas2021} it was shown that the effective dimension for QNNs tends to be larger than for CNNs, and from this it was argued that QNNs are generally more powerful. Furthermore, it was demonstrated that the encoding strategy has a significant impact on the effective dimension, and thus on the power, of quantum networks. In this work, we examine the effect that choice of QNN ansatz, and in particular choice of parameterised quantum gates, has on the effective dimension (and, by extension, the power of the network).

\begin{figure}
	\includegraphics[scale=1]{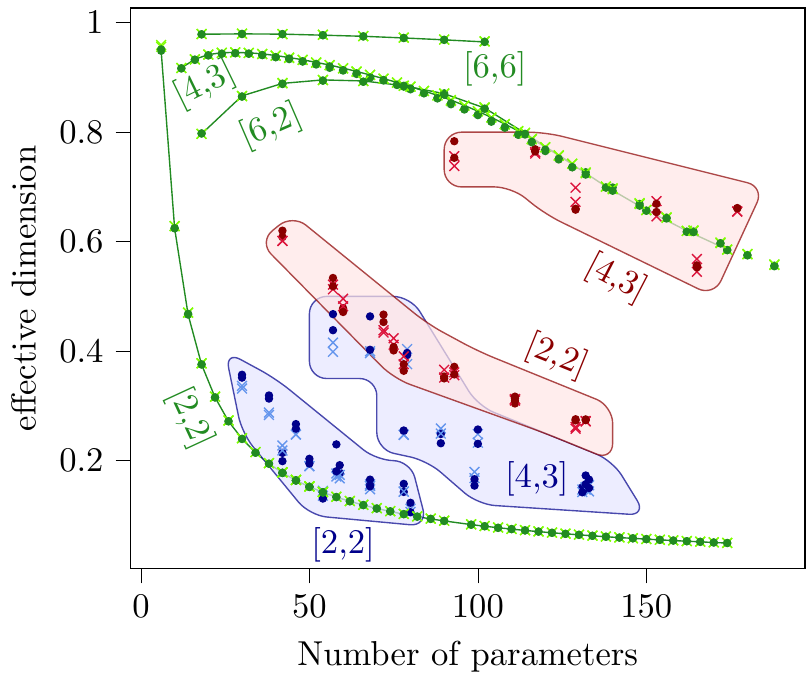}
	\centering
	\caption{\label{fig:EffectiveDimension} The effective dimension, normalised by the number of parameters, for different networks at $N_S = 10^4$ samples. SQP networks are shown in blue, Hannover networks in red and HEA in green. The number of input and output qubits are listed as $[N_\textrm{in},N_\textrm{out}]$. For a fixed number of input and output qubits, the number of parameters is varied by changing the number and width of the hidden layers. }
\end{figure}

In Fig.~\ref{fig:EffectiveDimension}, we see the effective dimension for various different networks as a function of the number of tunable parameters. In the plot the effective dimension is normalised by the total number of parameters. (One may note that in the limit that the number of samples $N_S \rightarrow \infty$, the effective dimension approaches the parameter dimension, $\textrm{dim}_{\textrm{eff}} \rightarrow N_\theta$). It can be seen that, for a given number of input and output qubits, the effective dimension of SQP networks is generally lower than that of the other networks studied. This indicates that, despite the UAT, SQP networks are not expressive compared with other QNN ans\"atze.

\subsection{Effective Capacity}\label{sec:EffectiveCapacity}

Another measure of the power of classical neural networks is the \textit{effective capacity} \cite{Zhang2016}. This is the ability of a learning model to fit randomized datasets. 
Thus, it is not concerned with generalisation or any true learning ability, but simply the ability of a network to fit raw data by force.
In part due to their universal approximation property, deep neural networks have been able to achieve near-zero training loss on completely randomised data.
This remarkable fact is part of the ``unreasonable'' effectiveness of deep learning \cite{Sun2017,Sejnowski2020}.

\begin{figure}
	\includegraphics[scale=1]{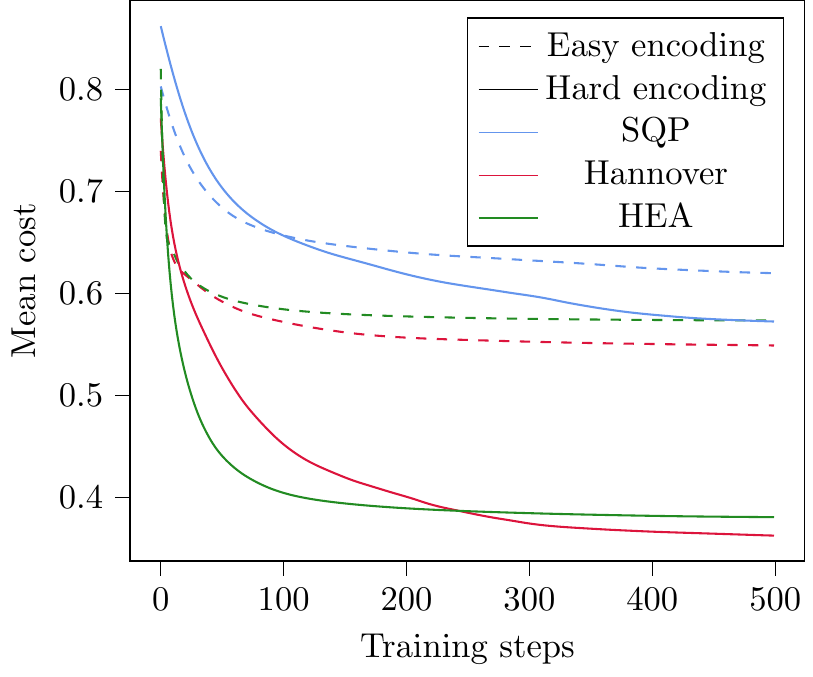}
	\centering
	\caption{\label{fig:RandomFit} Average training loss for various networks trained to fit a randomly generated data set. The the main text for a discussion of the different networks and different encoding strategies. Here we see that both encoding strategy and choice of network ansatz have a large impact on the ability of models to fit arbitrary functions. We also see that the SQP performs much more poorly than the UAT might have suggested.}
\end{figure}

In Fig.~\ref{fig:RandomFit} we train SQP, Hannover and HEA QNNS to fit completely random data -- that is, data where both training samples and corresponding labels are randomly generated.
The training is conducted using gradient descent via the Adam optimiser \cite{Kingma2014} with hyperparameters $\alpha=0.01$, $\beta_1=0.9$ and $\beta_2=0.99$.
Two different encoding schemes are employed: first, an ``easy'' feature map (so called because it can be efficiently simulated on a classical computer) which consists only of single qubit rotations applied to the input qubits (that is, angle-encoding), and a ``hard'' feature map introduced in \cite{Havlicek2019} which entangles the input data across multiple qubits (and as a result cannot be simulated efficiently on a classical computer).
It can bee seen that in all cases the ``hard'' feature map is better able to fit the data, indicating that this choice of encoding leads to a more expressive network.

For the QNNs we consider here, the number of quantum gates (and thus tunable parameters) is much to low too enjoy any of this unreasonable effectiveness.
However, the training loss achieved by our networks on randomized data still gives us a good indication of the power of a model in a context that is more indicative of realistic learning problems than the more general but more abstract measures like the effective dimension.
By examining the performance of a network on random data after training, we see not only how large the class of functions a network can fit is, but we also evaluate the ability of the network to find appropriate functions during training.

In general, we find that ranking models based on their ability to fit random data agrees well with the ranking based on their effective dimension.
Both choice of feature map and choice of gates play a role in determining the capacity of the network.
However we see that, regardless of choice of feature map, the SQP network does not perform as well at fitting random data as either the Hannover or HEA networks, and thus has a lower effective capacity.

\section{Numerical Experiments}\label{sec:Experiments}

Here we examine the performance of our QNNs on realistic learning tasks.

\subsection{Classification of real-world data sets}\label{sec:Classification}

A common task for artificial neural networks is the classification of some data set via supervised learning.
This has been a staple application for classical networks \cite{Goodfellow2016}, as well as a potential application for quantum neural networks.

As a concrete example, we take the well-studied iris dataset \cite{Dua2017} as our training data.
The input data for this dataset consists of four real numbers corresponding to the length and width of the petals and sepals of three distinct species of iris, and the task of the network is to correctly identify the flower based on these measurements.

\begin{figure}
	\includegraphics[scale=1]{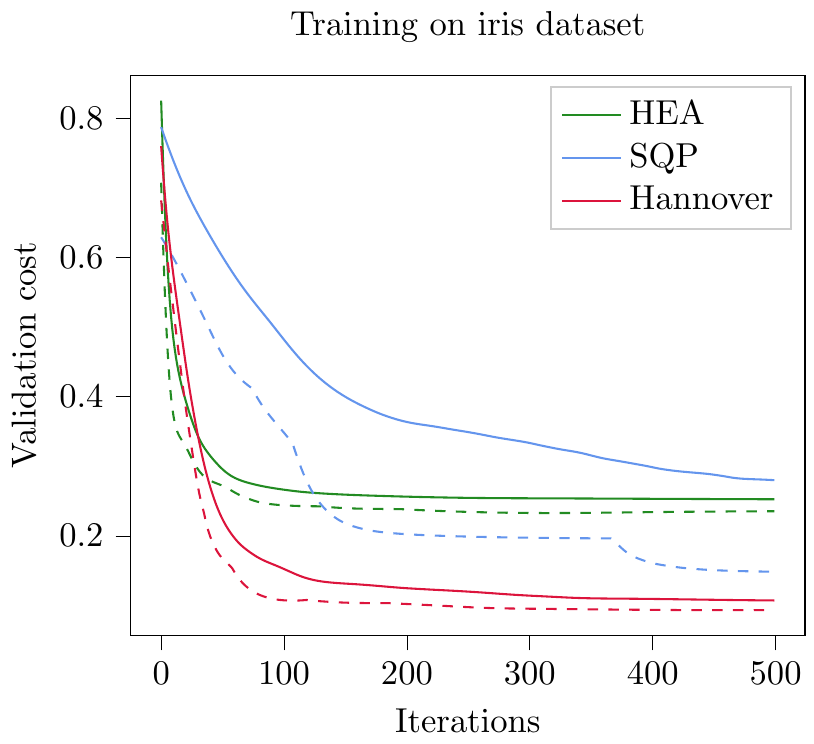}
	\centering
	\caption{\label{fig:Iris} Validation costs for QNNs trained on the \texttt{iris} dataset. Solid lines indicate average performance, dashed lines indicated minimum achieved cost. All networks use $N_S=30$ training samples. HEA has 32 hidden layers, DQNN has 2 hidden layers of width 3 and SQP has 2 hidden layers of width 4. }
\end{figure}

We train a network to classify the iris dataset using empirical risk minimisation with a least-squares cost function
\begin{equation}
C = \frac{1}{N_S} \sum_{j=1}^{N_S} | f_\theta(\textbf{x}_j) - y(\mathbf{x}_j) |^2
\end{equation}
where $\mathbf{x}$ is an input datapoint, $y(\mathbf{x})$ is the corresponding correct classification and $N_S$ is the number of training pairs. 
Training is conducted in the same manner as Sec.~\ref{sec:EffectiveCapacity}.
In Fig.~\ref{fig:Iris} we plot the cost that a network achieves on a previously unseen validation set of $N_V$ labelled pairs.
For each network, we try 16 different initialisations to circumvent the problem of getting stuck in local minima, and plot both the average validation cost and the lowest cost achieved.

Recent work has cast some doubt on the ability of QNNs to outperform classical networks on the classification of real-world classical data sets \cite{Qian2021}.
Here a QNN essentially serves as a special case of a parameterised model, and since the input data, output data \textit{and} training algorithm are all fundamentally classical, it is difficult to see where any room for genuine quantum advantage lies for such tasks.
A much more promising route for quantum machine learning is to take a situation where the input data is inherently quantum.
We consider such a situation in the next section.

\subsection{Unitary learning}\label{sec:UnitaryLearning}

A task more uniquely suited to quantum networks is learning a quantum map. That is, to find parameters $\boldsymbol \theta$ such that the circuit $\mathcal{U}(\boldsymbol{\theta})$ approximates some desired map $\mathcal{V} : \mathcal{H}^N \rightarrow \mathcal{H}^M$.

We restrict ourselves to considering only unitary maps and fix $N=M$. This is the situation that was considered in \cite{Beer2020} and implemented on hardware in \cite{Beer2021}. Here we can consider a target unitary $U_T$.
A network can be trained to approximate $U_T$ via supervised learning using training pairs consisting of input states $\ket{x}$ and output states generated by applying the target unitary to in the inputs $\ket{y(x)} = U_T \ket{x}$.
The network is trained via gradient ascent, maximising the objective function
\begin{equation} \label{eq:objective}
	O = \frac{1}{N_S}\sum_x \bra{y(x)}\rho_x(\boldsymbol{\theta})\ket{y(x)}
\end{equation}
where $\rho_x(\boldsymbol{\theta})$ is the output density matrix and $N_S$ is the number of training pairs.
For the HEA (or any other network in which the same set of qubits is used throughout), $\rho_x(\boldsymbol{\theta})$ is simply
\begin{equation}
\rho_x(\boldsymbol{\theta}) = \mathcal{U}(\boldsymbol{\theta})\ket{x}\bra{x}\mathcal{U}^\dagger(\boldsymbol{\theta}).
\end{equation}
For networks with a dissipative structure, such as the SQP and Hannover networks, the output density matrix must instead be considered the result of repeated applications of quantum channels
\begin{equation}
\rho_x(\boldsymbol{\theta}) = \mathcal{E}^\textrm{out}(\mathcal{E}^L(\dots \mathcal{E}^2(\mathcal{E}^1(\ket{x}\bra{x}))))
\end{equation}
with
\begin{equation}
	\mathcal{E}^{l(\rho^{l-1})} = \textrm{Tr}_{l-1}\left[\mathcal{U}^l(\boldsymbol{\theta}^l) \left(\rho^{l-1} \otimes \ket{0}^l\bra{0^l} \right)[{\mathcal{U}^l}^\dagger (\boldsymbol{\theta}^l) \right] 
\end{equation}
so that with each subsequent layer new qubits are added and the previous layers are traced out.
Note that here there is no need for a quantum feature map as out input data are already quantum states.

We consider two different classes of target unitaries: completely random unitaries, which are the result of choosing randomly generated complex numbers for the matrix elements of $U_T$, and Ising-generated random unitaries, where $U_T$ has the form
\begin{equation}
	U_T^\textrm{Ising} = \exp\left[-i\sum_{j}^{N} \left(J_j Z_j Z_{j+1} + h_j X_j \right)\right]
\end{equation}
with randomly chosen values for each $J_j$ and $X_j$, where $N$ is the number of qubits in the input and output layers.

\begin{figure}
	\includegraphics[scale=1]{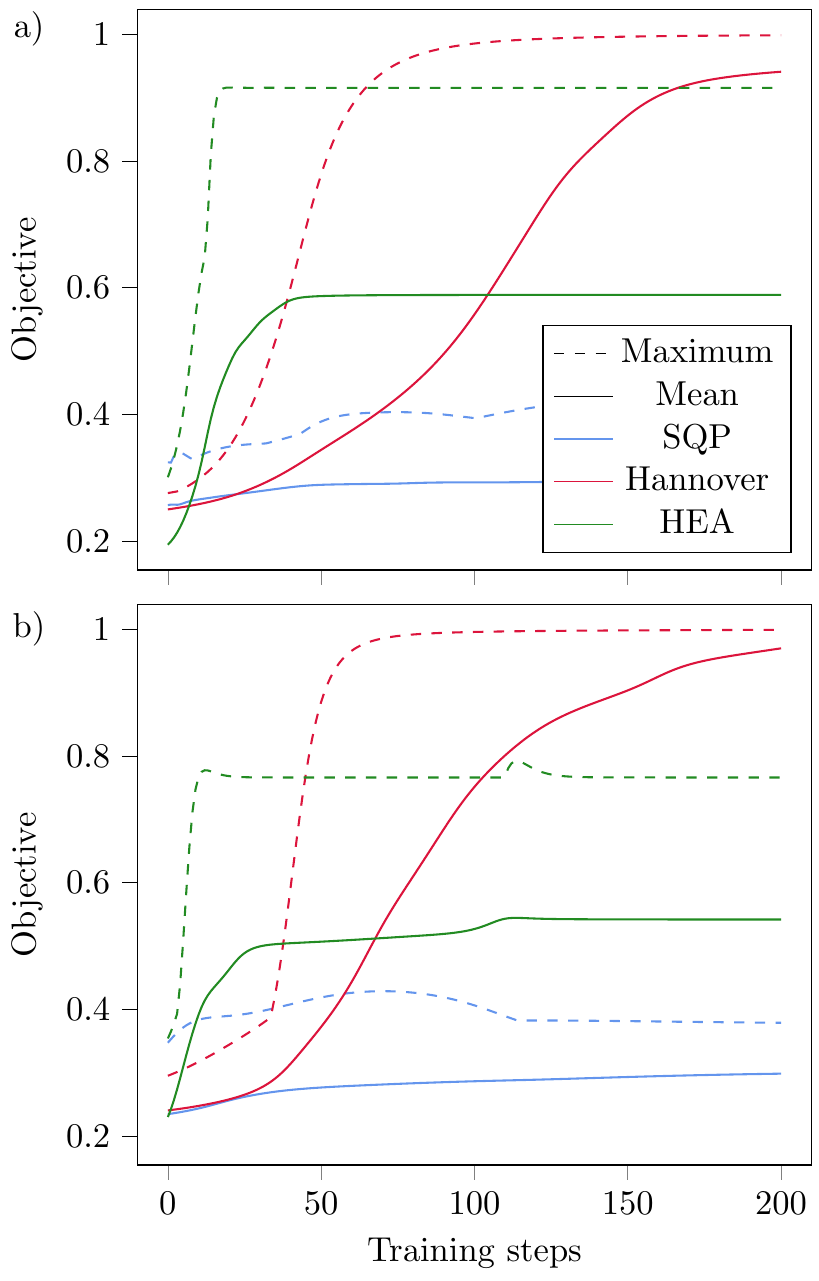}
	\centering
	\caption{\label{fig:UnitaryLearning} Maximum (dashed) and mean (solid) values of the objection function on the validation set when various networks are trained to reproduce randomly generated unitary operations. In a) the target unitary is generated by a transverse-field Ising model Hamiltonian with randomised couplings and fields, whereas in b) the target unitary is a more general random unitary matrix. All networks have two input and two output qubits. The HEA network has 32 hidden layers, the SQP has 3 hidden layers of width 3 while the Hannover network has 2 hidden layers of width 3. Learning rate $\eta=0.1$ for all HEA and Hannover networks, whereas the SQP network uses $\eta=5$ for a) and $\eta=2$ for b). 50 training pairs and 10 validation pairs are used throughout.}
\end{figure}

Both cases yield similar results, which can be seen in Fig.~\ref{fig:UnitaryLearning}.
There we show the performance of all three networks on unseen validation data, using parameters listed in the figure caption.
The Hannover network, which was designed for precisely this task, performs well, achieving a maximum value for the objective Eq.~\ref{eq:objective} of over 0.99 in both cases.
The HEA is less effective.
It reaches a maximum value very early on and then plateaus, with a large gap between maximum and mean values.
This indicates that the HEA network is very prone to getting stuck in local minimum, and thus is highly sensitive to the initialisation.

The SQP, on the other hand, shows very little improvement over the entire course of the training.
This network was unable to approximate the target unitary, even for a wide range of hyperparameters and intialisations.

\section{Conclusion}\label{sec:Conclusion}

SQPs are specialised parameterised quantum gates inspired by the sigmoid perceptrons of CNNs. They have been designed specifically to exhibit the universal approximation property that classical perceptrons enjoy, and it was hoped that this property would grant them enhanced expressive power when compared with more familiar parameterised quantum gates.

However, the numerical data presented in this paper does not support that hope. In fact, we have found that in many contexts simpler quantum circuits consisting only of standard gates outperform networks of SQPs. We examined three different QNN architectures: networks of SQPs, the feed-forward deep QNNs proposed by \cite{Beer2020} and simple circuits consisting only of single-qubit $Y$-rotations and CNOTs between pairs of qubits.

This work leads us to reconsider the way we think about the design of QNNs. Of the networks we have considered, the SQP is the one that best emulates a classical deep neural network. However, with only a limited number of qubits and gates available to us in the NISQ era, these networks cannot hope to approach the thousands or even millions of parameters enjoyed by classical deep networks. When restricted to small numbers of qubits ($\sim 1 - 10$) and small numbers of gates ($\sim 100$), these networks perform poorly at the tasks we have considered here. On the other hand, QNNs designed with a mind towards quantum hardware, rather than classical analogues, are able to perform well at both classical classification and the more inherently quantum task of unitary learning. 
The UAT, despite its importance to the theory of classical neural networks, should not serve as a guiding principle for the design on QNNs.

This work does not conclusively rule out the possibility of use-cases for SQPs.
However, based on the expressibility results we show that one should not generically expect SQPs to form powerful networks in the NISQ era.
Since SQPs have been shown to emulate classical neural networks, it is possible that they may form viable quantum machine learning models when it becomes feasible to build QNNs with tens of layers and millions of parameters, however such an assessment is beyond the scope of this manuscript.

\section*{Acknowledgements}

This work has received funding from the European Union's Horizon 2020 research and innovation programme under grant agreement No 828826 ``Quromorphic"

\appendix
\section{Derivation of adiabatic ramp unitary} \label{app:ARunitary}
In the original work of \cite{Torrontegui2019}, the adiabatic ramp (AR) quantum neuron is considered mostly as an input-output relation in the Heisenberg picture.
To easily describe this neuron protocol as a gate within a parameterised quantum circuit, we would wish to have a representation of this protocol as a unitary operation, so that the effect of successive applications of the gate, and its effect on arbitrary quantum states, is made clear.
Here we derive such a unitary from an assumption that the protocol proceeds exactly adiabatically.

The protocol begins with an Ising-type many-qubit Hamiltonian
\begin{equation}
	H = -\Omega(t)X_j + b_jZ_j - \sum_{k<j} w_{jk}Z_jZ_k.
\end{equation}
where the index $j$ labels the target qubit and the sum runs over every qubit in the previous layer, indexed by $k$.
It is assumed that $\Omega(t)$, $b_j$ and $w_{jk}$ are all parameters that can be freely chosen and tuned. The $w$ and $b$ parameters are the weights and biases of the network, which we wish to vary during training, while $\Omega(t)$ is a time-dependent field, which we will vary over the course of the operation of the gate.

Initially, $\Omega(t=0)$ is large so that the Hamiltonian for the target qubit is approximately
\begin{equation}
	H_\textrm{target}(t=0)\approx -\Omega(t=0)X_j
\end{equation}
so that the eigenstates involve the target qubit in a state $\ket{\pm}$.
The protocol calls for an initial Hadamard gate to be applied to the traget, such that if it was previously in a computational basis state it is brought into an $X$ eigenstate.
Let us describe the initial state of the multi-qubit system as
\begin{equation}
\ket{\psi(t=0)} = \mathop{\sum_{ z\in \{0,1\} }}_{  p\in\{\pm\}  }  ^{N_{L-1} } c_{z,p}\ket{z,p},
\end{equation}
so that we are expressing the state as a superposition of computational basis states $z$ of the input layer, and $X$ eigenstates $p$ of the target qubit. (Note that, since the only operators acting on the input layer are $Z$ operators, the computational basis states are also eigenstates at $t=0$).

Assuming that evolution is fully adiabatic, each initial eigenstate remains in an instantaneous eigenstate.
These eigenstates are given by 

\begin{equation} \label{eq:eigenstates}
	\ket{\phi_z^\pm} = \ket{z}\otimes\left[\sqrt{f[a(\mp z)/\Omega(t)]}\ket{0} \pm \sqrt{f[a(\pm z)/\Omega(t)]}\ket{1} \right]
\end{equation}
where
\begin{equation}
a_z = \sum_k w_{k} z_k - b_j 
\end{equation}
is sum of weights and biases and 
\begin{equation}
f(x) = \frac{1}{2}\left(1 + \frac{x}{\sqrt{1 + x^2}} \right)
\end{equation}
is the sigmoid activation function.
Note that $f(-x) = 1 - f(x)$.

$\Omega(t)$ is decreased adiabatically until some final $t=T$ at which point $\Omega=1$ and we have the final state
\begin{equation}
	\ket{\psi(t=T)} = \mathop{\sum_{ z\in \{0,1\} }}_{  p\in\{\pm\}  }  ^{N_{L-1} } c_{z,p} \ket{\phi_z^p(t_\textrm{final})}.
\end{equation}
We now know how the protocol maps every basis state, and can thus infer the unitary operation described in the main text.

\bibliography{Bib}

\end{document}